\documentclass[fleqn,11pt]{article}
\usepackage{amsmath,amssymb,amsfonts,graphicx}

\usepackage{amsfonts,amssymb,cite}

\def\noi{\noindent}

\newcommand{\Title}[1]{\noi {{\Large\bf #1}}\\[1ex]}

\def\Aunames#1{\noi{\bf #1}}
\def\au#1{${}^{#1}$}
\def\Addresses#1{\medskip\noi \protect
	\begin{description}\itemsep -3pt {\it #1} \end{description}}
\def\adr#1#2{\item[${}^{#1}$]{\it #2}}

\newcommand{\Abstract}[1]{\vskip 2mm \begin{center}
        \parbox{16.4cm}{\small\noi #1} \end{center}\medskip}

\def\email#1#2{\footnotetext[#1]{e-mail: #2}\addtocounter{footnote}{1}}

\def\nqq{\hspace*{-2em}}
\def\nhq{\hspace*{-0.5em}}

\def\cm{\hspace*{1cm}}
\def\inch{\hspace*{1in}}

\def\tall{\mbox{$\tst\vphantom{\int^0}$}}

\def\Jl#1#2{#1 {\bf #2},\ }

\def\ApJ#1 {\Jl{Astroph. J.}{#1}}
\def\CQG#1 {\Jl{Class. Quantum Grav.}{#1}}
\def\DAN#1 {\Jl{Dokl. AN SSSR}{#1}}
\def\GC#1 {\Jl{Grav. Cosmol.}{#1}}
\def\GRG#1 {\Jl{Gen. Rel. Grav.}{#1}}
\def\JETF#1 {\Jl{Zh. Eksp. Teor. Fiz.}{#1}}
\def\JETP#1 {\Jl{Sov. Phys. JETP}{#1}}
\def\JHEP#1 {\Jl{JHEP}{#1}}
\def\JMP#1 {\Jl{J. Math. Phys.}{#1}}
\def\NPB#1 {\Jl{Nucl. Phys. B}{#1}}
\def\NP#1 {\Jl{Nucl. Phys.}{#1}}
\def\PLA#1 {\Jl{Phys. Lett. A}{#1}}
\def\PLB#1 {\Jl{Phys. Lett. B}{#1}}
\def\PRD#1 {\Jl{Phys. Rev. D}{#1}}
\def\PRL#1 {\Jl{Phys. Rev. Lett.}{#1}}

\def\al{&\nhq}
\def\lal{&&\nqq {}}
\def\eq{Eq.\,}

\def\beq{\begin{equation}}
\def\eeq{\end{equation}}
\def\bear{\begin{eqnarray}}
\def\bearr{\begin{eqnarray} \lal}
\def\ear{\end{eqnarray}}
\def\earn{\nonumber \end{eqnarray}}

\def\nnn{\nonumber\\ \lal }
\def\nnnv{\nonumber\\[5pt] \lal }
\def\yy{\\[5pt] {}}
\def\yyy{\\[5pt] \lal }
\def\eql{\al =\al}

\def\dst{\displaystyle}
\def\tst{\textstyle}
\def\fracd#1#2{{\dst\frac{#1}{#2}}}

\def\Half{{\fracd{1}{2}}}

\def\e{{\,\rm e}}
\def\d{\partial}

\def\sign{\mathop{\rm sign}\nolimits}

\def\const{{\rm const}}

\def\then{\ \Rightarrow\ }
\newcommand{\toas}{\mathop {\ \longrightarrow\ }\limits }

\newcommand{\vars}[1]{\left\{\begin{array}{ll}#1\end{array}\right.}

\def\eqn#1{\eq\eqref{#1}}
\def\rf{\eqref}
\def\arccot{\mathop{\rm arccot}\nolimits}
\def\tall{\mbox{$\tst\vphantom{\int^0}$}}
\def\mn{_{\mu\nu}}
\def\MN{^{\mu\nu}}
\def\mN{_\mu^\nu}

\def\df{{\delta\phi}}
\def\ds{{\delta\psi}}

\def\cR{{\cal R}}

\def\kappa{\varkappa}

\def\hG{{\hat\Gamma}}
\def\N{{\mathbb N}}
\def\R{{\mathbb R}}
\def\oC{{\bar C}}
\def\og{{\bar g}}
\def\ophi{{\bar\phi}}
\def\oR{{\bar R}}

\def\ME{\mathbb{M}_{\rm E}}
\def\MJ{\mathbb{M}_{\rm J}}
\def\ssph{static, spherically symmetric}

\def\bh{black hole}
\def\bhs{black holes}
\def\wh{wormhole}
\def\whs{wormholes}
\def\asflat{asymptotically flat} 

\usepackage{color}

\begin{document}
\thispagestyle{empty}
\twocolumn[

\bigskip

\Title {Hybrid metric-Palatini gravity: black holes, wormholes, singularities\yy
	and instabilities}
	
\Aunames{K. A. Bronnikov,\au{a,b,c1} S. V. Bolokhov,\au{b;2} and M. V. Skvortsova\au{b;3}}

\Addresses{\small
\adr a {Center fo Gravitation and Fundamental Metrology, VNIIMS, 
		Ozyornaya ul. 46, Moscow 119361, Russia}
\adr b {Institute of Gravitation and Cosmology, RUDN University, 
		ul. Miklukho-Maklaya 6, Moscow 117198, Russia}
\adr c{National Research Nuclear University ``MEPhI'', 
		Kashirskoe sh. 31, Moscow 115409, Russia}
        }
                
\Abstract
   {The hybrid metric-Palatini theory of gravity (HMPG), proposed in 2012 by T. Harko et al., 
   is known to successfully describe both local (solar-system) and cosmological observations. 
   We discuss static, spherically symmetric vacuum solutions of HMPG with the aid of its 
   scalar-tensor representation. This scalar-tensor theory coincides with general relativity with a 
   conformally coupled scalar field (which can be canonical or phantom), therefore the known solutions 
   of this theory are re-interpreted in terms of HMPG. In particular, in the case of zero scalar field potential 
   $V (\phi)$, such that both Riemannian and Palatini Ricci scalars are zero, generic asymptotically 
   flat solutions either contain naked singularities or describe traversable wormholes, and there are 
   only special cases of black hole solutions with extremal horizons. There is also a one-parameter 
   family of solutions with an infinite number of extremal horizons between static regions.
   Examples of analytical solutions with nonzero potentials $V (\phi)$ are also described, among them 
   black hole solutions with simple horizons which are generic but, for canonical scalars, they require 
   (at least partly) negative potentials. With phantom scalars there are ``black universe'' solutions that 
   lead beyond the horizon to an expanding universe instead of a singularity. Most of the solutions under
   consideration turn out to be unstable under scalar monopole perturbations, but some special black 
   hole solutions are stable.
    }

\bigskip

] % -------------------------------------------------------------------------------------------------------------------
\email 1 {kb20@yandex.ru} 
\email 2 {boloh@rambler.ru}
\email 3 {milenas577@mail.ru}

% ===================
\section{Introduction}
% ===================

   General relativity (GR) that has recently celebrated its century, is known to still successfully 
   describe all local observational effects. It is however, unable to completely account for
   large-scale phenomena, facing the so-called Dark Matter and Dark Energy problems. 
   There are two alternative  ways in addressing these problems: one is still adhere to GR 
   but to introduce so far unobserved forms of matter like WIMPs (weakly interacting massive 
   particles) as Dark Matter, and a cosmological constant or a ``quintessence'' scalar field, etc., 
   as Dark Energy \cite{DE}. An alternative approach is to modify GR itself, considering  
   more general Lagrangian functions (for instance, $f(R)$), introducing new degrees of freedom 
   (e.g., scalar or vector fields), extra dimensions or/and geometric quantities such as torsion 
   and nonmetricity \cite{ex-GR1, ex-GR2}.

   The hybrid metric-Palatini gravity (HMPG) theory, proposed in \cite{har12}, is one of such theories.
   This theory assumes the existence of the Riemannian metric $g\mn$ along with an independent
   connection $\hG\mn^\alpha$. The total action reads \cite{har12}
\beq  				\label{S}
	S = \frac{1}{2\kappa^2}\int d^4 x\sqrt{-g} [R + F(\cR)] + S_m,
\eeq    
   where $R = R[g]$ is the scalar curvature derived from $g\mn$, while $F(\cR)$ is a function
   of the scalar $\cR = g\MN \cR\mn$ obtained with the Ricci tensor $\cR\mn$ built in the 
   standard manner from the connection $\hG\mn^\alpha$; also, $g = \det(g\mn)$, 
   $ \kappa^2$ is the gravitational constant, and $S_m$ is the action of nongravitational matter.
  
   Thus HMPG combines the metric and Palatini approaches to gravity and is an extension of
   $f(R)$ theories. This theory has been shown to agree with the classical gravitational tests
   in the Solar system \cite{cap13b}, rather well describes the dynamic properties of galaxies 
   and galaxy clusters, thus approaching an explanation of the dark matter problem \cite{cap13a}, 
   and is able to create models of the accelerating Universe without a cosmological constant,
   see reviews \cite{cap15,har18} for a more detailed description of HMPG and 
   its achievements. A further generalization of HMPG, with an arbitrary function of both $R$ 
   and $\cR$, is suggested in \cite{boh13}, see recent results obtained in this theory in 
   \cite{mont19, lem19, lem20}.
  
   The present paper continues the study of \ssph\ solutions of HMPG, began in \cite{kb19},
   where the simplest case $F(\cR) \propto \cR$ was considered. In this case, the HMPG theory 
   is equivalent to GR with a conformally invariant scalar field that can be either canonical or 
   phantom; the phantom case, which seems to appear more naturally from HMPG
   (since then $dF/d\cR > 0$), was discussed in \cite{kb19}. Here we briefly reproduce the results 
   of \cite{kb19}, add a discussion for the canonical $\phi$ field, and also present two simple
   analytically solvable cases of fields with nonzero potentials that correspond to more complex
   $F(\cR)$ than $F \sim \cR$. 
       
   The paper is organized as follows. The next section discusses the basic features of the
   STT representation of HMPG \cite{har12,cap15}. Sec. 3 is devoted to \ssph\ solutions in 
   the massless case ($V(\phi) =0$) for both canonical and phantom $\phi$ fields. 
   Sec. 4 discusses analytical solutions with $V(\phi) \ne 0$, also with canonical and phantom 
   fields. In all cases we pay special attention to globally regular solutions 
   and solutions containing Killing horizons, in particular, possible black hole solutions with 
   nonzero potentials. Section 5 is a brief consideration of the stability of all HMPG solutions 
   discussed in this paper under radial (monopole) perturbations. Section 6 contains
   some concluding remarks.
  
% =========================================================
\section {Basic features of HMPG and its scalar-tensor representation}  
% =========================================================
  
   Variation of \rf{S} with respect to the independent connection $\hG\mn^\alpha$ leads 
   to the conclusion \cite{cap15, dan19} that $\hG\mn^\alpha$ is the Levi-Civita connection
   corresponding to a metric conformal to $g\mn$, namely $h\mn = \phi g\mn$, with the 
   conformal factor $\phi = F_\cR \equiv dF/d\cR$. It shows that this theory actually contains, 
   in addition to $g\mn$, only one dynamic degree of freedom expressed in the scalar field
   $\phi$. As shown in \cite{har12,cap15}, the whole theory admits a reformulation as a 
   scalar-tensor theory with the gravitational part of the action  
\beq                           \label{S1} 
	S_g = \! \int \! d^4x \sqrt{-g}\bigg[(1+\phi)R - \frac {3}{2\phi}(\d\phi)^2 - V(\phi)\bigg], 
\eeq          
   where\footnote
  	{Unlike \cite{har12, cap15, dan19} etc., we are using the metric signature $(+  -   -\, -)$, 
  	hence the plus sign before $(\d\phi)^2 = g\MN \phi_{\mu}\phi_{\nu}$ corresponds to a 
  	canonical field and a minus to a phantom field. We will also safely omit the factor 
  	$1/(2\kappa^2)$ at the gravitational part of the action since only vacuum configurations, 
  	where $S_m =0$, will be considered. The Ricci tensor is defined as 
  	$R\mn =\d_\nu  \Gamma^\alpha_{\mu\alpha} - \ldots$, so that, for example, the scalar 
  	curvature is positive in de Sitter space-time. We also use the units in which $c = G = 1$
  	(c being the speed of light and $G$ the Newtonian gravitational constant.} 
   the potential $V(\phi) $ is related to $f(\cR)$ by
\beq                             \label{VR}
		V(\phi) = \cR F_\cR - F(\cR).
\eeq    
 
   The theory with the action \rf{S1} evidently belongs to the Bergmann-Wagoner-Nordtvedt 
   class of STT \cite{berg68, wag70, nor70} in which the gravitational action is
\beq   	 	\label{S-STT}
	S_g =\int\! d^4x \sqrt{-g}\Big[f(\phi)R  + h(\phi)(\d\phi)^2 - V(\phi)\Big],
\eeq     
   with arbitrary functions $f(\phi)$, $h(\phi)$ and $V(\phi)$. In our case, $V$ is given 
   by \rf{VR}, while
\beq  			\label{f,h-I}
	f(\phi) = 1+\phi, \cm     h(\phi) = - \frac{3}{2\phi}. 	
\eeq 
    
   The general action \rf{S-STT} admit a well-known transformation \cite{wag70}
   to the Einstein conformal frame in which the scalar field is minimally coupled to the 
   metric (while the formulation \rf{S-STT} is called the Jordan conformal frame). 
   The transformation reads \cite{wag70}
\bearr 			\label{J-E}
	\og\mn = f(\phi) g\mn, \qquad  \frac {d\phi}{d\ophi} = f (\phi) | D(\phi)|^{-1/2},
\nnn	
	D(\phi) = f(\phi)h(\phi) + \frac 32 \bigg(\frac{df}{d\phi}\bigg)^2,
\ear
  and results in 
\beq  	\nhq		\label{S-E}
	S_g = \int \! d^4x \sqrt{-\og}
		\bigg[\oR + n \og\MN \ophi_{,\mu}\ophi_{,\nu} - \frac{V(\phi)}{f^2(\phi)}\bigg],
\eeq                      
   where bars mark quantities obtained from or with the transformed metric $\og\mn$,
   while the factor $n = \sign D(\phi)$ distinguishes canonical scalar fields ($n =+1$) with 
   positive kinetic energy from so-called phantom fields ($n =-1$) with negative kinetic 
   energy.
  
   In the theory \rf{S1} we have $D = -3/(2\phi)$ and $n = - \sign \phi$, so that 
\bear      \nhq \label{can}
	\phi = -\tanh^2 \frac{\ophi}{\sqrt 6} && (n = +1,\ \  -1 < \phi <0)  ,
\\                  \label{pha}
	\phi = \tan^2 \frac{\ophi}{\sqrt 6} \quad\  &&   (n = -1, \ \  \phi > 0). 
\ear    
   Thus, depending on the sign of $\phi$, the theory splits into canonical and phantom 
   sectors, and the emergence of the latter looks more natural since in this case all three
   metrics $g\mn$, $\og\mn$ and $h\mn = \phi g\mn$ have the same signature. Let us 
   also note that values of $\phi$ smaller than $-1$ lead to a negative effective gravitational 
   constant and are thus manifestly nonphysical.
  
   The substitution $\phi = - n \chi^2/6$ converts the action \rf{S1} to the form
\bearr                                     \label{S2}                          
		S_g = \int \! d^4x \sqrt{-g} \Big[(1 - n \chi^2/6)R 
\nnn \inch	\cm	
		+ n (\d\chi)^2 - W(\chi)\Big],
\ear
  with $W(\chi) = V(\phi)$. The action \rf{S2} describes GR where the source of gravity 
  is a conformally coupled scalar field (as mentioned in \cite{cap15}), which has the usual 
  sign of kinetic energy if $\phi < 0$ ($n =1$) and is of phantom nature if $\phi > 0$ ($n = -1$). 
  Conformally coupled scalar fields have been considered in a great number of studies, 
  beginning with those of Penrose \cite{pen64} (a massless conformally invariant field) and 
  Chernikov and Tagirov \cite{tag68} (massive conformally coupled fields). The theory \rf{S2}  
  with a phantom scalar was also discussed in  \cite{ZK} as a possible alternative to GR in 
  astrophysical and cosmological applications. 
   
  In the massless case, $V(\phi) \equiv W(\chi) =0$, the field equations due to \rf{S1} or \rf{S2}
  imply that all vacuum solutions (such that $S_m=0$) have both zero Ricci scalars, 
  $R = \cR = 0$ (see \cite{dan19}). A general inverse result is also valid \cite{kb19}:
   
 \medskip 
   {\it Let there be a vacuum solution with $R \equiv 0$ and a non-constant scalar field in 
   a theory \rf{S-STT}, with $V \equiv 0$, then this STT reduces either to GR with a conformally 
  coupled scalar field (which may be canonical or phantom) or to pure conformal scalar 
  field theory.} 
       
\medskip       
  The transition \rf{J-E} is well known as a method of finding exact or approximate solutions 
  to the field equations due to \rf{S-STT} since the equations due to \rf{S-E} are simpler 
  than those due to \rf{S-STT}. An Einstein-frame solution having been found, its Jordan-frame 
  counterpart is easily produced by a transformation inverse to \rf{J-E}.
  
  There is, however, an important subtle point: if the function $f(\phi)$ in \rf{S-STT} turns to zero
  or infinity at some value of $\phi$, it may happen that a singularity in the Einstein-frame manifold 
  $\ME$ with the metric $\og\mn$ transforms into a regular surface in the Jordan-frame manifold 
  $\MJ$ with the metric $g\mn$ (or vice versa), and $\MJ$ should then be continued beyond this 
  surface. Such a phenomenon, termed {\it conformal continuation} \cite{kb01p,kb02}, has been 
  observed in  special cases of a number of scalar-vacuum and scalar-electrovacuum solutions, 
  in particular,  those of GR with conformally coupled scalar fields \cite{kb70, kb73} and in the 
  Brans-Dicke theory \cite{CBH1, CBH2} (the so-called cold black holes).  

  All \ssph\ solutions with $V \equiv 0$ are well known, but since they admit a new interpretation 
  in terms of HMPG, it makes sense to discuss them from this viewpoint, it is done in Section 3.
  A large number of scalar-vacuum solutions with $V \not\equiv 0$ are also known
  (see, e.g. \cite{shik02, pha1, pha2, zlo04, anab12} and references therein), 
  and we will discuss some of them in the context of HMPG in Section 4. 
     
   A question of interest is: suppose we have found a solution of STT with some $V(\phi)$,
   then, what is the corresponding HMPG? In other words, given $V(\phi)$, can we determine
   $F(\cR)$? 
   
   For the case $V(\phi) \equiv 0$, \eqn{VR} gives simply $F(\cR) = \const\cdot \cR$.  
   For $V(\phi) \not\equiv 0$, since $\phi = F_\cR$, the relation \rf{VR} is a 
   Clairaut equation (see, e.g., \cite{kamke}) whose solution consists of a regular family 
   that contains only linear functions,
\beq         \label{cle1}
		F(\cR) = H\cR - V(H), \qquad H = \const,
\eeq      
  and the so-called singular solution which is an envelope of the regular family
  and may be presented in a parametric form:
\bearr                       \label{cle2}
		F(\cR) = \phi\cR - V(\phi),
\nnn
		\cR = dV/d\phi.
\ear            
   This issue is discussed in more detail in \cite{dan19}.
     
% ==============================
\section{Solutions for $V(\phi) \equiv 0$}
% ==============================

   In the case $V(\phi) \equiv 0$, solutions to the Einstein-minimally coupled scalar equations 
   can be written in a unified form for canonical and phantom scalars using the harmonic 
   coordinate condition \cite{kb73}
\beq         \label{harm}
		\alpha(u) = 2\beta(u) + \gamma(u),
\eeq    
   in terms of the general \ssph\ metric in $\ME$
\bearr                                \label{ds_E}
	  ds^2_{\rm E} = \e^{2\gamma}dt^2 - \e^{2\alpha}du^2 - \e^{2\beta} d\Omega^2,
\nnn  \inch
	  d\Omega^2 = d\theta^2 + \sin^2 \theta\, d \varphi^2.
\ear
   The solution reads 
\bearr                  \label{sol-E}
           \ophi = \oC u + \ophi_0,\qquad \gamma(u) = - hu, 
\nnn    \nhq       
           e^{-\beta(u)-\gamma(u)} = s(k,u) := \vars {\!
                        k^{-1}\sinh ku,  \ & k > 0 \\
                                    u,  \ & k = 0 \\
                        k^{-1}\sin ku,   \ & k < 0,  }           
\nnn
	   h, \ k,\ \oC,\ \ophi_0 = \const,
\ear           
  where, without loss of generality, the radial coordinate $u$ is defined at $u > 0$ ($u=0$ 
  corresponds to flat spatial infinity),  while the integration constants $h$, $k$ and $\oC$ 
  (the scalar charge) are constrained by the relation
\beq                                                       \label{int-0}
            2k^2\sign k = 2h^2 + n \oC^2,
\eeq
  where, as before, $n=+1$ corresponds to a canonical field and Fisher's solution \cite{fish48},
  and $n =-1$ to its phantom counterpart \cite{b-lei57, h_ell73} (sometimes called the 
  ``anti-Fisher'' solution). The metric \rf{ds_E} now reads
\beq 			\label{ds-Fish}
            ds_E^2 = \e^{-2hu} dt^2 - \frac {\e^{2hu}}{s^2(k,u)}
            					\bigg[\frac {du^2}{s^2(k,u)} + d\Omega^2 \bigg].
\eeq      
  As follows from \rf{int-0}, with $n=+1$ we have $k >0$, hence there 
  is a single branch, whereas for a phantom scalar the solution splits into three branches
  according to \rf{sol-E}, with qualitatively different properties. Their detailed descriptions 
  may be found, e.g., in \cite{bbook12, kb-stab11}.

% ------------------------------------------------------  		  		
\subsection{The canonical sector}
% ------------------------------------------------------  	

  According to the above-said, the Jordan-frame metric and the scalar field $\phi$ in the 
  theory \rf{S1} with $V(\phi)=0$ and $\phi < 0$ ($n=+1$) may be presented as \cite{kb73}
\bearr                   \label{ds_J1}
             ds_J^2 = \cosh^2 \psi \bigg\{ \e^{-2hu}dt^2 
\nnn \cm     
             - \frac{k^2 \e^{2hu}}{\sinh^2 ku} \bigg[\frac {k^2 du^2}{\sinh^2 ku} + d\Omega^2 \bigg]\bigg\},
\yyy                       \label{phi1}
  	     \phi (u) = \tanh^2 \psi,  \qquad   \psi := \ophi/\sqrt{6} = Cu + \psi_0,     
\ear      
  where the notation  $\psi = \ophi/\sqrt{6}$ has been introduced for convenience, $C = \oC/\sqrt{6}$,   
  $\psi_0 = \ophi_0/\sqrt{6}$, and the constant $k$ is expressed via $C$  and $h$:
\beq 			\label{int1}
		k = \sqrt {h^2 + 3 C^2}.
\eeq      
  The solution is defined at $u > 0$ so that $u=0$ corresponds to spatial infinity, near it, the 
  spherical radius $r =\sqrt{ -g_{22}}$ behaves as $r \sim 1/u$, and the Schwarzschild mass 
  is\footnote
  		{If we write the general static, spherically symmetric metric in the form \rf{ds_E} with 
  		  an arbitrary radial coordinate $u$, this metric is asymptotically flat at some 
  		  $u = u_0$ if \cite{bbook12}
 	\[
 		     \e^{\beta(u)} \equiv r(u) \toas_{u\to u_0} \infty, \quad |\gamma(u_0)| < \infty,
 		     		\quad  \e^{\beta -\alpha}|\beta'|  \toas_{u\to  u_0} 1.
	\] 		  
		Then, comparing \rf{ds_E} with the Schwarzschild metric, it is easy to obtain a general 
		expression for the Schwarzschild mass at $u = u_0$ :
	\[
		m = \lim\limits_{u \to u_0} \e^\beta \gamma'/\beta'.
	\]
		In particular, for the (anti-)Fisher metric \rf{ds-Fish} we have $m = h$ at $u_0 = 0$.
		}  		  
\beq   \label{m1}
  		m_J =  h - C \tanh \psi_0.
\eeq  
   At the other end of the $u$ range, as $u \to \infty$, there are three kinds of behavior:
\begin{itemize}     
	\item
		$C < h$:\ we have $g_{00} \to 0$, and $r \to \infty$. It is a naked attracting singularity 
		located beyond a throat, the kind of singularity called a ``space pocket'' by 
		P. Jordan \cite{jord}.
	\item
		$C > h$: \ in this case, $g_{00} \to \infty$ and $r \to 0$, so this is a naked singularity 
		at the center, repulsive for test particles.
	\item
		$C =h >0$: both $g_{00}$ and $r$ tend to finite limits, so that $u = \infty$
		is a regular sphere, and a continuation beyond it is necessary. 
\end{itemize}
    To extend the solution for $C=h$ beyond $u = \infty$, let us put 
\beq                      \label{uy} 
    y = \coth hu, \qquad   u = \frac 1{2h} \ln \frac {y+1}{y-1}.
\eeq    
    The metric becomes
\bearr                    \label{ds_J1a}
	   ds_J^2  = \frac{(y+y_1)^2}{1-y_1^2}\bigg[\frac{dt^2}{(y+1)^2} 
\nnn \cm\cm	   
	   			- \frac{h^2}{y^4} (y+1)^2 (dy^2 + y^2 d\Omega^2)\bigg],
\ear
  where $y_1 = \coth (\psi_0/C)$. The sphere $u=\infty\ \leftrightarrow\ y =1$ is now manifestly 
  regular.  We obtain:
\begin{itemize}     
\item
	$y \to \infty$ is flat spatial infinity. 
\item
	$y_1 <  0 \ \then  \ y = - y_1 > 0$ is a {\it naked attracting singularity} at the center ($r\to 0$).
\item
	$y_1 >  0 \  \then  \ y \to 0$ is one more flat infinity, and the whole configuration is a 
	{\it traversable \wh}.
\item
	$y_1 =  0  \then  y =0$ is a double horizon. Passing on to the coordinate $r = h(y+1)$, 
	we obtain
\beq            \nqq         \label{ds_J1b}
	ds_J^2 = \Big(1\! - \!\frac hr \Big)^2 dt^2 - \Big(1\! - \!\frac hr \Big)^{-2} dr^2 - r^2 d\Omega^2,
\eeq	
  which is the well-known {\it \bh\ solution with a scalar charge and a conformal 
  scalar field} \cite {kb70, bek74}, sometimes called the BBMB \bh\  solution.	
\end{itemize}
  One can notice that the substitution \rf{uy} loses its meaning at $y < 1$. Accordingly,
  the relation \rf{can}, that is, $\phi = - \tanh^2 \psi$, is also meaningless at $\phi < -1$.
  Instead, after the conformal continuation, we have \cite{kb02} in the Einstein frame     
  another ''copy'' of the Fisher solution, where, instead of \rf{can}, 
  $\phi = - \coth^2 \psi$, and the conformal factor in \rf{ds_J1} is $\sinh^2 \psi$. 
  At the transition surface $y =1$, the field  $\phi$ crosses the critical value $\phi = -1$,
  and beyond it, at $ \phi < -1$, there is an ``antigravitational'' region, with a negative effective 
  gravitational constant, where, in other words, the graviton becomes a ghost \cite{b-star07}.  

  We see that the solutions with $V \equiv 0$ and $n=+1$ generically contain naked 
  singularities, while the only existing \bh\ and \wh\ solutions are special, emerge due to 
  conformal continuations, and each of them contains an ``antigravitational''  region.

% ---------------------------------------------------
\subsection{The phantom sector}
% ---------------------------------------------------
 
  Assuming $\phi > 0, \ n = -1$, we obtain, quite similarly to \rf{ds_J1},
\bearr                   \label{ds_J2}
             ds_J^2 = \cos^2 \psi \bigg\{ \e^{-2hu}dt^2 
\nnn \cm     \cm       
             - \frac{e^{2hu}}{s^2(k,u)}\bigg[\frac {du^2}{s^2(k,u)} + d\Omega^2 \bigg]\bigg\},
\yyy                       \label{phi2}
  	     \phi (u) = \tan^2 \psi,  \qquad   \psi := \ophi/\sqrt{6} = Cu + \psi_0,     
\ear      
  For the integration constants $k, h$ and $C$ we now have
\beq 			\label{int2}
		k^2 \sign k = h^2 - 3 C^2.
\eeq      
  We see that $\sign k$ is not fixed, and accordingly the solution splits into three branches.

  Let us assume, without loss of generality, $|\psi_0| < \pi/2$. Then, as before, the metric 
  \rf{ds_J2} is \asflat\ at $u=0$ ($r \equiv \sqrt{-g_{22}} \to \infty$ where
  $r \sim 1/u$,\footnote 
  		{The conformal factor $\cos^2 \psi$ is not normalized to unity at $u=0$
  		if $\psi_0 \ne 0$, which, however, does not affect the further description.}
 and the Schwarzschild mass is
\beq  			\label{m2}  
  		m = h + C \tan \psi_0.
\eeq  
  Other properties of the solution depend on the sign of $k$, taking into account
  the definition of $s(k,u)$ in \rf{sol-E}.   
  
\medskip\noi
  {\bf Branch A:} $k > 0$.The metric reads
\bearr                   \label{ds_A}
             ds_J^2 = \cos^2\psi \bigg\{ \e^{-2hu}dt^2 
\nnn \inch
                 - \frac{k^2 \e^{2hu}}{\sinh^2 ku} 
                 		\bigg[\frac {k^2 du^2}{\sinh^2 ku} + d\Omega^2 \bigg]\bigg\},      
\nnn
                      \psi = \psi_0 + Cu. \cm h^2 = 3C^2 + k^2.				                 		
\ear             
   The only difference from \rf{ds_J1} is the conformal factor $ \cos^2 (Cu + \psi_0)$ instead 
   of $ \cosh^2 (Cu + \psi_0)$, which drastically changes the metric behavior. Indeed, as  
   $u$ grows from zero, $\psi(u)$ ultimately reaches the value where $\cos\psi =0$ where, 
   according to \eqn{pha}, $\phi \to \infty$. This happens where 
   $Cu + \psi_0 \to \pi/2$ if $C > 0$ and where $Cu + \psi_0 \to -\pi/2$ if $C < 0$. Other 
   quantities involved in the metric are there evidently finite. Thus it is a naked central 
   (since $r \to 0$)  singularity, and it is attractive for test particles due to $g_{00} \to 0$.   
   
\medskip\noi
  {\bf Branch B:} $k = 0$. In this case, the solution has the form
\bearr                     \label{ds_B}
	  ds_J^2 = \cos^2\psi \bigg[ \e^{-2hu}dt^2 - \e^{2hu}
	  			\bigg(\frac{du^2}{u^4} + \frac {d\Omega^2}{u^2}\bigg) \bigg],
\nnn	  
     		\psi = \psi_0 + Cu, \cm h^2 = 3C^2.
\ear               
   As in Branch A, the coordinate $u$ ranges from zero  to the value where $\cos\psi =0$ 
   (say, $\psi = \pi/2$) and $\phi = \infty$, and we observe a central attractive singularity. 

\medskip\noi
  {\bf Branch C:} $k < 0$. Now the solution reads
\bearr              \nhq              \label{ds_C}
	ds_J^2 =  \cos^2\psi \bigg[\e^{-2hu}dt^2 - \frac{k^2 \e^{2hu}}{\sin^2 ku}
\nnn \inch  \cm    \times	
	                        \bigg(\frac{k^2 du^2}{\sin^2 ku} + d\Omega^2\bigg) \bigg],
\nnn 
                      \psi = \psi_0 + Cu, \cm h^2 = 3C^2 - k^2.				
\ear    
   The solution behavior crucially depends on $\psi_0$ at given $k$, $C$ and depends on 
   which of the quantities $\sin |k|u$ or $\cos\psi$ will be the first to vanish as $u$ grows 
   beginning from zero. For asymptotic flatness we should assume that at $u=0$ the factor 
   $\cos^2\psi$ is nonzero, hence $|\psi_0| < \pi/2$ without loss of generality. Then, three 
   possible behaviors should be singled out, see Fig.\,1 (we assume for certainty $C > 0$). 
% -------------------------------------------------------------------------   fig 1
\begin{figure}   
\centering
\includegraphics[width=6cm]{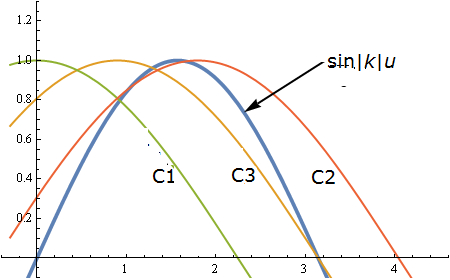}  
\caption{\small
	The behaviors C1, C2, C3 of the metric \rf{ds_C} is illustrated by the corresponding curves. 
	We assume $|k| = 1$; curves C1--C3 plot $\cos\psi$   for $C = 0.7$ and different $\psi_0$.
	Curve C1 corresponds to a naked singularity, C2 to a wormhole, and C3 to a black hole.. }       
\end{figure}   
% -------------------------------------------------------------------------

\medskip\noi
  {\bf C1:} $(\pi/2 - \psi_0)/C < \pi/|k|$. The solution terminates at $u = u_s = (\pi/2 - \psi_0)/C$, 
  where $\cos\psi =0$, and $u=u_s$ is a naked central singularity quite similar to the 
  one in branches A and B. 

\medskip\noi
  {\bf C2:} $(\pi/2 - \psi_0)/C > \pi/|k|$. The solution terminates at $u_*=\pi/|k|$
  where $\sin ku =0$, corresponding to the second flat spatial infinity, where the radius $r$ 
  infinitely grows while $g_{tt}$ and $\phi$ remain finite, and the Schwarzschild mass is there
  equal to
\beq      \label{m*}
                   m_* = -\e^{hu_*}(h \cos \psi_* + C \sin \psi_*),
\eeq    
  ($\psi_* = \psi_0 + C\pi/|k| < \pi/2$ is the value of $\psi$ at $u=u_*$). Such a \wh\ solution
  is only quantitatively different from its anti-Fisher and Brans-Dicke analogs,
  see, e.g., \cite{kb73,brss10}. 
   
\medskip\noi   
  {\bf C3:}  $(\pi/2 - \psi_0)/C = \pi/|k|$. In this intermediate case, at $u= u_1= \pi/|k|$ vanish 
  both $\sin |k|u$ and $\cos\psi$ , the spherical radius $r = \sqrt{-g_{\theta\theta}}$ is finite 
  but $\phi = \infty$. Near $u = u_1$, the metric behaves as
\bearr                                 \label{ds_hor}
		ds_J^2 = C^2 \bigg[\! \e^{-2hu_1} \Delta u^2 dt^2
			- \! \e^{2h u_1}\frac{du^2}{\Delta u^2} 
\nnn \inch \inch		
			-  \! \e^{2h u_1} d\Omega^2\bigg],
\ear      
  where $\Delta u = u_1 - u$. Consequently, $u=u_1$ is a double (extremal) horizon,
  and the metric should be continued beyond it. 
  
  The condition $(\pi/2 - \psi_0)/C = \pi/|k|$ leads to 	$\psi_0 = \pi (1/2 - C/|k|)$, thus 
  $C < |k|$, hence the plot of $\cos \psi$ is wider than that of $\sin ku$. Therefore, as $u$ 
  further grows (describing the region beyond the horizon), the next zero of $\sin ku$, 
  ($u = u_2 = 2\pi/|k|$) is reached before a zero of $\cos\psi$. The value $u = u_2$ 
  corresponds to the second flat spatial infinity, in full similarity with \wh\ solutions. 
  This space-time is also globally regular but is now only one-side traversable due to 
  emergence of the horizon. Figure 2 shows the corresponding Carter-Penrose diagram. 
  
% -------------------------------------------------------------------------   fig 2
\begin{figure}   
\centering
\includegraphics[width=4.5cm]{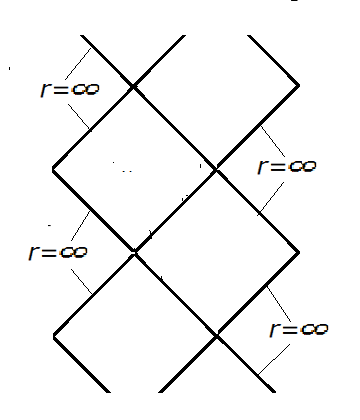}   :
\caption{\small
	Carter-Penrose diagram for a regular \bh\ with the metric \rf{ds_C}.  
	The diagram infinitely extends up and down.     }
\end{figure}   
% -------------------------------------------------------------------------
    
  This \bh\ solution has much in common with the one with the metric \rf{ds_J1b}
   \cite{kb70, kb73, bek74}. Both \bhs\ are described by special solutions to the 
   Einstein-scalar equations, both are \asflat\ and extremal (with zero Hawking temperature),
   and in both cases the supporting scalar fields turn to infinity on the horizon, whereas 
   the effective stress-energy tensors $T\mN$ are finite there (as is evident from finiteness 
  of the Einstein tensor components $G\mN$). Also, in both cases the scalar curvature is 
  zero in the whole space, and  the solutions are obtained  from their Einstein-frame 
  counterparts using conformal continuations \cite{kb73, kb02}.
  
  However, the solution \rf{ds_J1b} has a singular center $r=0$ (the geometry is the same as 
  that of the extreme Reissner-Nordstr\"om space-time), while the space-time \rf{ds_C}
  is globally regular and has no center at all. 

  It happens that none of the \ssph\ solutions of the theory \rf{S1} with $V \equiv 0$ have
  simple horizons with finite Hawking temperature, which contradicts the results announced in 
  \cite{dan19}.
\bigskip            
            
\medskip\noi
{\bf A geometry with infinitely many horizons.}           
  There is a one-parameter family of solutions of interest obtained if we put 
\beq                            \label{C-inf}
		C = |k|, \qquad \psi_0 = -\pi/2,
\eeq    
  (thus abandoning the asymptotic flatness requirement). We then have 
  so that $\cos^2 \psi = \sin^2 ku$, and the Jordan-frame metric reads
\beq  \nhq                     \label{ds_inf}
		ds_J^2 = \sin^2 ku \e^{-2hu} dt^2 - k^2 \e^{2hu}
				\bigg( \frac{k^2 du^2}{\sin^2 ku} + d\Omega^2 \bigg),
\eeq    
  where $h = \pm k$ according to \rf{int2}. This metric, with $u \in \R$, describes 
  a space-time unifying an infinite number of static regions (each described by a half-wave 
  of the function $\sin ku$), separated by double horizons located at each $u = \pi n/|k|$, with 
  any integer $n$. In this case, the Jordan-frame manifold $\MJ$ unifies a countable number 
  of Einstein-frame manifolds $\ME$, each of the latter representing an anti-Fisher \wh\ whose
  both infinities turn into horizons in $\MJ$. Another example of a manifold obtained by 
  infinitely many conformal continuations was obtained in \cite{kb02}, using a solution for a 
  conformally coupled scalar field $\phi$ with a nonzero potential $U(\phi)$ and the normal 
  sign of kinetic energy.  In that example, the transition from one region to another occurred 
  through ordinary surfaces $S_{\rm trans}$ of finite radius, there were no horizons, 
  and the whole $\MJ$ was either completely static (shaped as an infinitely long tube with a 
  periodically changing radius) or completely cosmological (forming a (2+1) cosmology 
  with a periodically changing scale factor). In the present case, all transitions surfaces 
  $S_{\rm trans}$ are double horizons, and the structure is aperiodic due to the factors
  $\e^{\pm hu}$ in \rf{ds_inf}. The corresponding global structure is shown in Fig.\,3.
  
% -------------------------------------------------------------------------   fig 3
\begin{figure}   
\centering
\includegraphics[width=4.5cm]{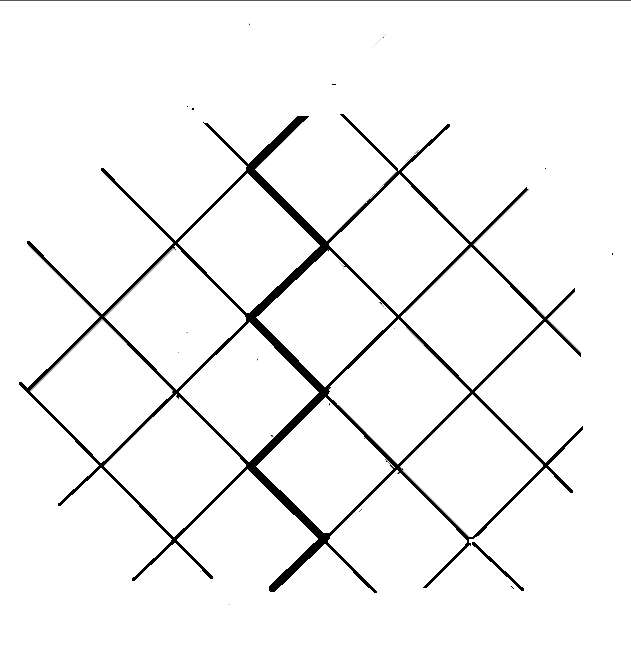}  
\caption{\small
	         Carter-Penrose diagram of the manifold $\MJ$ with the metric \rf{ds_inf}. The 
	         diagram occupies the whole plane. The thick broken line shows a horizon
	         corresponding to a particular value of $u = n\pi/|k|$, $n \in \N$. 
	}
\end{figure}   
% -------------------------------------------------------------------------  

  Another example of a manifold with infinitely many horizons \cite{cle09} has been obtained 
  for a family of phantom dilaton-Einstein-Maxwell \bhs.
  
% ==============================================
 \section{Solutions for $V(\phi)\not\equiv 0$}     
% ==============================================
 
  Before discussing particular examples, it makes sense to recall some general theorems
  concerning scalar-vacuum space-times with $V(\phi) \not\equiv 0$. Owing to such theorems, 
  one can say much about the possible behavior of solutions to the field equations even 
  being unable to solve them analytically. Many of these results concern the properties of 
  minimally coupled scalar fields, for example, no-hair theorems (see, e.g, \cite{herd15} 
  for a recent review) indicating the conditions that exclude the existence of horizons, 
  and global structure theorems \cite{kb01} telling us about possible regular solutions 
  and a maximum possible number of horizons. In particular, if $V(\phi) \geq 0$, an \asflat\ 
  \bh\ with a nontrivial canonical scalar field is impossible \cite{ad-pier}. On the other hand,   
  by \cite{kb01}, spherically symmetric scalar-vacuum space-times cannot contain more 
  than two horizons, and this number is only one for \asflat\ configurations. This result holds 
  for both canonical and phantom scalars with any $V(\phi)$. 

  For Jordan-frame space-times, conformal to those with minimally coupled scalar fields,
  most of the theorems are preserved without changes if the conformal factor $f(\phi)$ in \rf{J-E} 
  is everywhere finite and regular since (at transitions to either side) a flat infinity maps to a flat 
  infinity, a horizon maps to a horizon, and the potential $V(\phi)$ preserves its sign. The situation
  changes if $f(\phi)$ is somewhere zero or infinite, then the mapping can change the nature of 
  singularities, if any, and conformal continuations can emerge. The above examples show  that 
  such continuations can be numerous, up to an infinite number, as we saw in the Jordan-frame 
  metric \rf{ds_inf}. In particular, the number and nature of horizons in $\MJ$ may be different from
  that in $\ME$, including the possible number of simple horizons. It is, however, important 
  that conformal continuations can only emerge at special values of integration constants \cite{kb02}.
  
  As in the above massless case ($V \equiv 0$), HMPG solutions with nonzero potentials split 
  into the canonical and phantom sectors, in which the conformal factors $1/f(\phi) = 1/(1 +\phi)$ 
  have the forms $\cosh^2 \psi$ and $\cos^2\psi$, respectively. The first one is able to blow up
  and the second one to vanish, so in both cases the nature of solutions in $\MJ$ can be quite 
  different from that in $\ME$. We will briefly analyze the behavior of such solutions in some 
  particular cases of $V(\phi)$ admitting analytical solutions, known from the literature 
  \cite{shik02, pha1, zlo04}. 
  
% ------------------------------------------------------------- 
\subsection{$V \not\equiv 0$, the canonical sector}    
% -------------------------------------------------------------     
      
{\bf Example 1} . This solution (in the Einstein frame) has been obtained by the inverse problem 
  method \cite{shik02}  for a minimally coupled scalar field in the metric
\beq       \label{ds-q}
	ds_E^2 = A(x) dt^2 - \frac{dx^2}{A(x)}  - r^2 (x) d\Omega^2,  
\eeq
  (that is, \rf{ds_E} under the so-called quasiglobal coordinate condition $\alpha+\gamma=0$)
  by assuming 	
\beq                        \label{rE1}
	r(x) = \sqrt{x^2 - a^2}, 
\eeq
  where $a$ plays the role of a length scale. Let us assume $a=1$, thus expressing all quantities 
  in terms of this arbitrary length scale. Since one of the Einstein equations for the action \rf{S-E} 
  reads 
\beq                              \label{EE01}  
  		2r''/r = -n \ophi'{}^2, 
\eeq  
  and since now $r''/r = - (x^2-1)^{-2} < 0$, we are dealing with a 
  canonical scalar field, $n=+1$. Using as before, $\psi = \ophi/\sqrt{6}$, we obtain the 
  \asflat\ (as $x\to \infty$) solution in the form  \cite{shik02}  	
\bear                             \label{AE1}
	A(x) \eql 1 - 3mx + \frac 32 (x^2-1) \ln \frac{x+1}{x-1},	
\yy           			\label{psiE1}
	\psi (x) \eql \frac 1{2\sqrt{3}} \ln \frac{x+1}{x-1} + \psi_0, \quad\ \psi_0 = \const,  
\yy					\label{UE1}
	U(\psi) \eql \frac{3m}{x^2-1} \bigg[ 6x + (3x-1)\ln \frac{x+1}{x-1}\bigg], 
\ear         
  where $U(\psi)$ is the Einstein-frame potential according to \rf{S-E}:
\beq                  \label{U}
              U(\psi) = V(\phi) /(1 + \phi)^2  
\eeq      
  (in \rf{UE1} it is a function of $x$, but its $\psi$ dependence is easily restored by substituting 
  $x = x (\psi)$ determined from \rf{psiE1}).
  In this solution, $x\to \infty$ is flat spatial infinity, $ m $ is the Schwarzschild mass, and the value
  $x = 1$ corresponds to the spherical radius  $r=0$, which is a naked central singularity if
  $m \leq 1/3$ and a singularity hidden under an event horizon if $m > 1/3$, see Fig.\,4.
% ------------------------------------------------ fig 4   
\begin{figure}
\centering
\includegraphics[width=7cm]{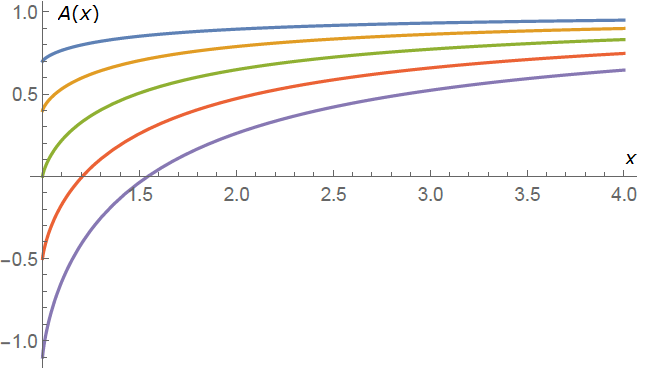}          
\caption{\small
   The function $A(x)$ according to \rf{AE1}  for $m = 0.1, 0.2$, $1/3, 0.5, 0.7$ (upside down).  It has a simple zero if $m > 1/3$, 
   the solution then describes a black hole. All solutions have a singularity at $x=-1$, where $r=0$. }
\end{figure}
% ------------------------------------------------ 
   In the ``massless''  case, $m = 0$, we have $A \equiv 1$, $U \equiv 0$, and the present solution 
   coincides with the case $h = 0$ of the solution \rf{ds_J1}, \rf{phi1}, though expressed using another
   radial coordinate.

  The potential $U$ is proportional to $m$, it is everywhere negative, singular at $x=1$
   and rapidly vanishes at infinity:
\bearr         \label{UE1-as}
	U(x) \sim \frac{\ln (x-1)}{x-1} \ \ {\rm as} \ \ x\to 1,
\nnn	
	U(x) \approx -\frac{8m}{5 x^5}\ \qquad {\rm as} \ \ x\to \infty.	
\ear

  The  Jordan-frame metric is 
\beq                      
     ds_J^2 = \cosh^2 \psi \  ds_E^2.
\eeq     		
     The conformal factor $\cosh^2 \psi$ is well-behaved at $x >1$,  tends to a constant at large $x$ 
     (so that the metric $ds^2_J$ is also asymptotically flat), and blows up as $x \to 1$: 
\bearr
       		\cosh^2 \psi \approx \cosh^2 \psi_0 + \frac{\sinh(2 \psi_0)}{\sqrt{3} x}, \quad\   x \to \infty,
\nnn       
                 \cosh^2 \psi \sim (x-1)^{-1/\sqrt 3}, \qquad  x \to 1.
\ear  
    Meanwhile, in the same limit $x \to 1$, 
\[ 
	    A(x) \approx 1-3m +3m(x-1)[\ln 2 -1 - \ln(x-1)].   
\]	    
    Thus the conformal factor cannot regularize the metric at $x=1$:
    it enhances the singularity if $m\neq 1/3$ (for example, $g_{tt}$ remains finite in the Einstein
    frame but blows up in Jordan's) and only modifies it if $m=1/3$. As a whole, the conformal 
    factor only deforms the metric at $x>1$ but does not change it qualitatively. We conclude
    that the HMPG solution with the potential according to \rf{U}, that is,
\beq
		V (\phi) = U(\psi) f^2(\phi) = U(\psi)/ \cosh^4  \psi
\eeq             
     describes a black hole with a simple horizon in the case $m > 1/3$. The black hole mass
     is equal to $m$ in the Einstein frame,  while in Jordan's we have
\beq               \label{m-E1}
                    m_J = m \cosh \psi_0 - \frac 1 {\sqrt{3}} \sinh \psi_0.
\eeq        
   They coincide if $\psi_0 =0$.
    
% -------------------------------------------------------------------------    
\noi
{\bf Example 2}. Another Einstein-frame solution with the metric \rf{ds-q} has been obtained 
  in \cite{zlo04} with the so-called separability approach but can also be found in full similarity 
  with \rf{rE1}--\rf{UE1} by assuming 	
\beq                           \label{rE2}
	r(x) = \sqrt{x (x+a)}, 
\eeq
  where $a$ again plays the role of a length scale and can be put equal to unity.
  The solution now reads
\bear                          \label{AE2}
		  A(x) \eql 1 - 6m (2x+1) 
\nnn \cm\ \   		  
		  + 12 mx (x+1) \ln \frac {x{+}1}{x},
\\                                \label{psiE2} 
		  \psi(x) \eql \Half \ln \frac {x+1}{x}  + \psi_0,
\yy                             \label{UE2} 		  
                  U(\psi) \eql  -\frac {12m}{x(x+1)}\Big[-3 (1 + 2 x) 
\nnn \cm  \ \               
		                  + (1 + 6 x + 6 x^2) \ln \frac{x+1}{x}\Big].                  
\ear
   The properties of this solution are quite similar to those of \rf{rE1}--\rf{UE1}. The canonical nature
   of the scalar field is assured by the fact that $r''/r = - 1/[4x^2(x+1)^2] < 0$. The solution has 
   a naked singularity at $x=0$ if $m \leq 1/6$ and describes a \bh\ with a simple horizon if
   $m > 1/6$, and the plots of $A(x)$ for different $m$ look almost the same as in Fig.\,4.
   The conformal factor $\cosh^2 \psi$ deforms the metric but does not remove the singulatities.

% -------------------------------------------------------------                                                                   
\subsection{$V \not\equiv 0$, the phantom sector}    
% -------------------------------------------------------------     

  Consider an analytic solution for a minimally coupled phantom scalar $\psi$ \cite{pha1}, 
  also obtained by the inverse problem method. Now we assume 
\beq                           \label{rE3}
	r(x) = \sqrt{x^2 +a^2}, 
\eeq
  and, as before, put $a=1$ as an arbitrary length scale. The inequality $r''/r = (x^2+1)^{-2} > 0$ 
  confirms  the phantom nature of $\psi$ and hence the original scalar $\phi$. Then, with the 
  Einstein-frame metric \rf{ds-q}, we have the solution \cite{pha1}
\bearr    	                     \label{AE3}
	  A(x) = 1 + 3mx - 3m (x^2+ 1) \arccot x ,	
\yyy                            \label{psiE3}
	\psi (x) = \frac 1{\sqrt{3}} \arctan x + \psi_0,   
\yyy				        \label{UE3}
	U(\psi) = \frac{6m}{(x^2+1)} \big[ -3x + (3x^2+1) \arccot x \big]. 
\ear         
  In this solution, $x \in \R$, $x\to \infty$ is flat infinity, and $m$ has the meaning of the Schwarzschild 
  mass. The behavior of the solution as $x \to -\infty$ is different, depending on the sign of $m$:
% ---------------------------------------------------------------------- fig 5   
\begin{figure*}    
\centering
\includegraphics[width=7cm]{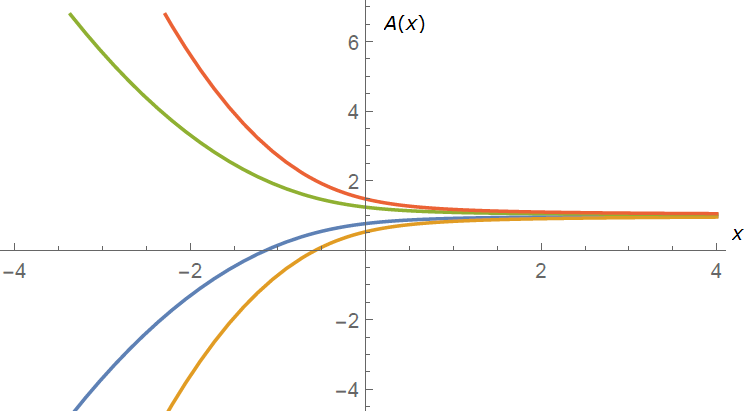}          
\qquad
\includegraphics[width=7cm]{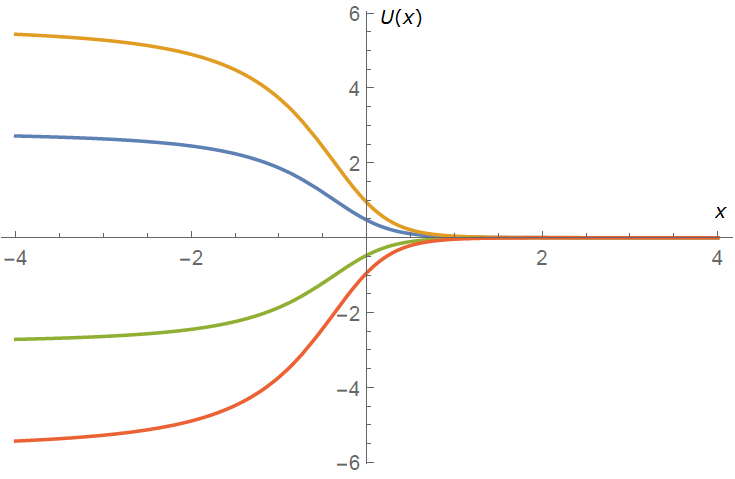}          
\caption{\small
         The metric function $A(x)$ (left panel) and the potential $U(\phi(x))$ (right panel)
         for the Einstein-frame solution \rf{ds-q}, \rf{AE3}--\rf{UE3}, $m = -0.1, -0.05, 0.05, 0.1$
         (bottom-up for $A(x)$, upside down for $U(x)$. As $x \to -\infty$, the metric is 
         asymptotically de Sitter if $m >0$ and AdS if $m < 0$.}
\end{figure*}    
% ---------------------------------------------------------------------- 
\medskip
\begin{itemize}
     \item
     		$m < 0:$\  $A \sim x^2,\ U \to \const < 0$ --- the solution describes a wormhole with an AdS 
     		limit at the ``far end''.
     \item
     		$m=0:$  \  $A \to 1,\ U \equiv 0$ --- it is the simplest (Ellis) twice \asflat\ \wh\ with 
     		zero mass \cite{kb73, h_ell73} ($A \equiv 1$, $U \equiv 0$).	  
     \item
     		$m > 0:$ \ $A \sim -x^2, \ U \to \const >0$ --- we obtain a regular \bh\ with a de Sitter 
     		expansion far beyond the horizon instead of a singularity	(a ``{\it black universe}'' 
     		\cite{pha1, pha2}). 
\end{itemize} 
   The behavior of $A(x)$ and $U(x)$ is shown in Fig.\,5.

  In Jordan's frame we have the metric 
\beq                             \label{ds_J3}
     	ds_J^2 = \cos^2 \psi  \Big[A(x) dt^2 - \frac {dx^2}{A(x)} - r^2(x) d\Omega^2\Big],
\eeq  
  and the geometry crucially depends on the value of $\psi_0$.  The range of $\psi$ is 
\beq              \label{range}
     {\rm Range}\ (\psi) = \Big(\psi_0 - \dfrac {\pi}{2\sqrt{3}},\ \psi_0 + \dfrac {\pi}{2\sqrt{3}}\Big),
\eeq
  its length is $\pi/\sqrt 3 < \pi$, smaller than $\pi$, length of the segment where $\cos\psi > 0$. 
  The spatial asymptotic value $x \to \infty$ corresponds to  
  $\psi = \psi_1 = \psi_0 +\pi/(2\sqrt{3})$, and the Schwarzschild mass is there
\beq
		m_J = m \cos \psi_1 - \frac 1{\sqrt{3}} \sin \psi_1. 
\eeq      
  If $\cos\psi \ne 0$ in the whole range \rf{range} (for example, if 
  $|\psi_0| < \pi (\sqrt 3 -1)/(2\sqrt 3) \approx 0.663$), then the conformal factor $\cos^2 \psi $ is 
  everywhere positive and regular, it then only deforms the metric $ds_E^2$ but does not change 
  it qualitatively.

 Otherwise,  the conformal factor $\cos^2 \psi$, in general, creates a singularity at some finite 
 $x = x_s$. It destroys wormhole solutions, producing a central attracting singularity instead of their
 regular far end; a similar singularity is created instead of a horizon in black universe solutions 
 if $x=x_s$ belongs to a static region or is precisely a horizon; it is like a big bang (or crunch)
 if it happens to be in a nonstatic region beyond the horizon. Lastly, if $\psi_0 = \pi/2 - \pi(2\sqrt 3)$, 
 so that  $\psi_1 = \pi/2$, then the conformal factor destroys the asymptotic flatness of the solution.
 
 We conclude that in the phantom sector all three kinds of solutions are generic: \bh\ ones,\wh\ 
 ones and those with naked singularities. The \bh\ solutions can be regular (black-universe type)
 or singular beyond the horizon, depending on the value of $\psi_0$. 
 
% ==========================
\section{Stability}
% ==========================

  Since the transition \rf{J-E} to the Einstein frame may be viewed as simply a change of variables 
  in the differential equations, it can be applied to the perturbed field equations on equal grounds 
  with those for static configurations. This enables us to use the existing results of the studies of 
  small perturbations of scalar-vacuum space-times with minimally coupled scalar fields, see, e.g., 
  \cite{b-kh79,hay02,ggs1,ggs2,kb-stab11,kb-stab12,kor15,kor17,chap17}. 
  The Jordan frame perturbations obey the same equations as in the Einstein frame, but only 
  expressed in other variables. However, the Jordan-frame stability inferences may be different 
  since the boundary conditions should now be formulated according to the physical requirements 
  inherent to $\MJ$. 
 
  Let us discuss the stability properties of the solutions described above with respect to 
  purely radial (monopole) perturbations. The experience indicates that these perturbations 
  are, in a clear sense, the most dangerous for configurations with scalar fields: if a system 
  is unstable, it is most probably a monopole mode that implements this instability.    
  A physical reason for that is that the effective potentials for all other perturbations contain
  centrifugal barriers which are positive and therefore favorable for stability. 
  
% -------------------------------------------------
\subsection{Perturbation equations}
% -------------------------------------------------  

\def\eff{_{\rm eff}}
     
  It is well known that in \ssph\ scalar-vacuum space-times, monopole perturbations of the 
  whole system are governed by the scalar field perturbations $\delta\phi(u,t)$, or those 
  of the Einstein-frame field, $\delta\psi(u,t)$, representing the only dynamic degree of 
  freedom. These perturbations obey a single linear equation whose coefficients depend 
  of the parameters of the background static system, while the metric perturbations 
  $\delta\alpha, \delta\beta, \delta\gamma$ (in terms of the metric \rf{ds_E}) can be found from 
  the solutions of the ``master equation'' for $\delta\psi$. The master equation for a spectral 
  component of the perturbation, $\delta\psi = \Psi(u) \e^{i\omega t}$, in the Schr\"odinger-like 
  canonical form,  
\beq                    \label{Schr}
		\frac{d^2 Y}{dz^2}  + \big(\omega^2 - W\eff(z)\big) Y =0.
\eeq    
  In this equation, $z$ is the so-called tortoise radial coordinate such that 
  $du/dz = \e^{\gamma-\alpha}$, where $u$ is an arbitrary radial coordinate in the metric
  \rf{ds_E}. The unknown function in \rf{Schr} is $Y(z) = \Psi(u) e^\beta$, while the effective 
  potential $W(z)$ has the form  \cite{kb-stab11,kor15,kor17}
\bearr   \nhq   \label{Weff}
		 W\eff(z)\! = \!\e^{2\gamma} \bigg[\frac{3n\psi'^2}{\beta'^2}(U\! - \! 2\e^{-2\beta})
			   \!+\! \frac{\psi'}{\beta'} U_\psi \!+\! \frac n{12} U_{\psi\psi}	  \bigg]  
\nnnv		\cm  
		  + \e^{2\gamma-2\alpha}[\beta'' + \beta'(\beta'+\gamma'-\alpha')],
\ear 
  where the index $\psi$ denotes $d/d\psi$, the prime denotes $d/du$ ($u$ is again an 
  arbitrary coordinate in \rf{ds_E}), and $U = U(\psi) = V(\phi)/(1+\phi)^2$, see \rf{U}.
  It should be stressed here that the notations $\alpha, \beta, \gamma$ 
  refer to the metric \rf{ds_E} written in the Einstein frame.
  
  Solving this equation with appropriate boundary conditions, we find a spectrum of 
  eigenvalues $\omega^2$ of this boundary-value problem, and, as usual, if there are 
  $\omega^2 < 0$, we can conclude that the background configuration is unstable under 
  linear monopole perturbations since there is a time-dependent perturbation growing as 
  $\e^{|\omega|t}$. To assert that the instability is inherent to the configuration itself rather 
  than caused by energy pumping from outside, it is also necessary to verify that there is 
  no energy flow into the system through the boundaries. However, this requirement does
  not lead to any new restrictions for our system: indeed, quite similarly to the 
  reasoning in \cite{b-kh79}, at flat infinity the energy flux is zero for any admissible 
  solution to \rf{Schr}, while at the other end the flow direction is controlled by the arbitrary
  sign  in a solution to \rf{Schr} and can always be chosen so that the energy leaks outward.
  
  We will discuss the stability properties of the configurations enumerated above, using 
  as much as possible the previous results available in the literature.
  
  At flat spatial infinity, where both fields $\phi$ and $\psi$ are regular, we naturally require
  both $\df \to 0$ and $\ds \to 0$. In what follows we assume that this requirement is always
  applied, and focus on boundary conditions on the other end of the range of the radial 
  coordinate.  
      
% -------------------------------------------------------
\subsection{Stability: the canonical sector}
% -------------------------------------------------------  
 
  {\bf 1.} $V(\phi) \equiv 0$, solution \rf{ds_J1}, \rf{phi1}, the conformally mapped Fisher 
  solution in the general case ($C \ne h$). At the singularity $u \to \infty$ we have $\phi \to -1$, 
  and due to its fixed value it is natural to require $\df \to 0$ for meaningful perturbations. 
  Since $\phi = - \tanh^2 \psi$, for $\psi$ this requirement transforms to 
  $\ds = o(\e^{2\psi})$, where $\psi\to \infty$,
  whereas the instability of Fisher's solution was established under the much weaker 
  boundary condition $|\ds/\psi| < \infty$ \cite{b-kh79}. 
  The present condition $\ds = o(\e^{2\psi})$ is much weaker, therefore, the perturbations 
  which grow with time in Fisher's solution, thus implementing its instability, manifestly 
  satisfy the new, weaker boundary conditions, and we conclude that
  the solution \rf{ds_J1}--\rf{int1}  is unstable.
  % ------------------------------------------------ fig 6  
\begin{figure}
\centering
\includegraphics[width=8.5cm]{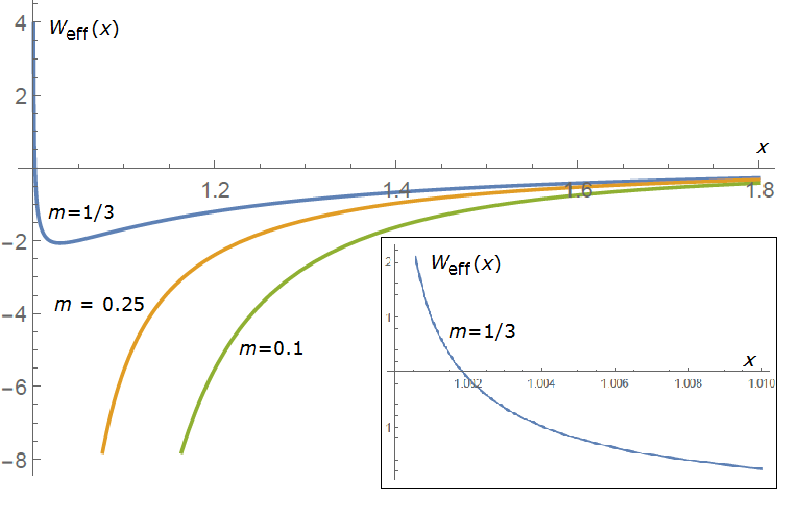}          
\caption{\small
   The potential $W\eff (x)$ for the solution (39)--(41) with naked singularities, $m = 0.1, 0.25, 1/3$. 
   The inset shows a more detailed behavior of $W\eff (x)$ for $m =1/3$ near $x = 1$. }
\end{figure}
% ------------------------------------------------ 
  
\medskip\noi  
  {\bf 2.} $V(\phi) \equiv 0$, solution \rf{ds_J1a}, $y_0>0$ (a wormhole). 
  This solution is unstable as proved in \cite{gri01, gri05}. 
  The instability is related to the existence of a negative pole $W\eff (z) \approx -1/(4z^2)$
  at the transition sphere $z = 0$ ($u = 1$) of the conformal continuation. This
  leads to the existence of negative eigenvalues $\omega^2$ of
  the corresponding boundary-value problem.

\medskip\noi
  {\bf 3.} $V(\phi) \equiv 0$, solution \rf{ds_J1a}, $y_0< 0$ (a naked singularity beyond
  the transition surface of conformal continuation).The instability conclusion follows from
  the same reasoning as in the previous case.
  
\medskip\noi
  {\bf 4.} $V(\phi) \equiv 0$, solution \rf{ds_J1b} (black hole). This solution is stable as 
  proved in  \cite{turok}, although previously \cite{kb78} the opposite result was announced.
% ------------------------------------------------ fig 7 
\begin{figure*}
\centering
\includegraphics[width=5.7cm]{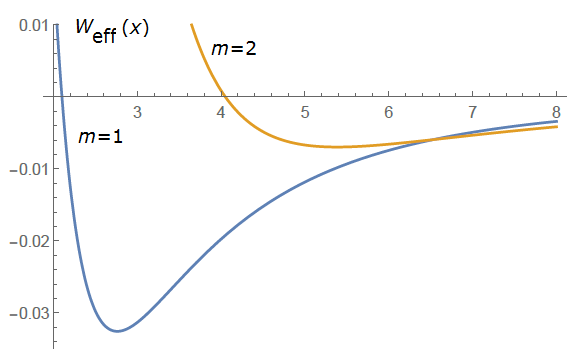}          
\includegraphics[width=5.7cm]{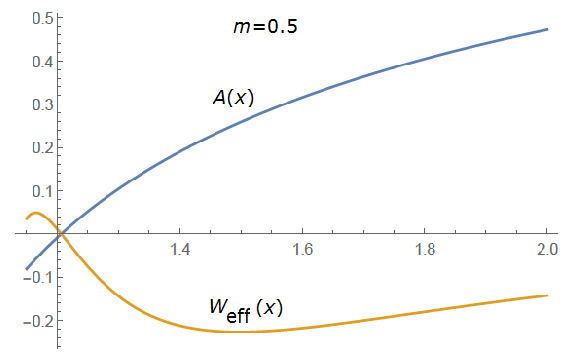}          
\includegraphics[width=5.7cm]{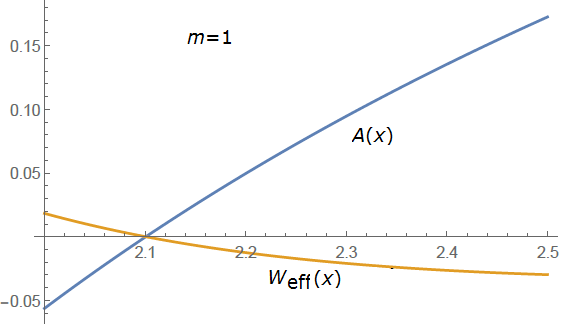}          
\caption{\small
   Generic behavior of $W\eff (x)$ for the \bh\ solution (39)--(41) $m > 1/3$ (left panel),
   and comparison of $W\eff (x)$ with  $A(x)$ (middle and right panels).}
\end{figure*}
% ------------------------------------------------       

\medskip\noi
  {\bf 5.} $V(\phi) \not\equiv 0$, solution \rf{ds-q}, \rf{AE1}--\rf{UE1}, $m \leq 1/3$ 
  (a naked singularity). For this case, the shape of the effective potential (60) is shown 
  in Fig.\,6, and it asymptotically behaves as follows:
\bearr
	W\eff= - \frac{2m}{x^3} + \frac{-3+4m^2}{x^4} + O(x^{-5}),
\nnn \inch	
  		x\to \infty,
\yyy  		
	W\eff= - \frac {(3m\!- \!1)^3}{4(x-1)^2} + \frac{27m^2\! -\! 3}{4(x-1)} + O\big(\ln^2(x-1)\big),
\nnn \inch
		x\to 1.	
\ear      
  Thus for $m < 1/3$ the potential $W\eff  \to - \infty$, but $W\eff  \sim \ln(x-1)^2$ for $m = 1/3$.
  On the other hand, the ``tortoise'' coordinate $z = \int dx/A(x)$ is found as follows for $x \to 1$:
\bearr
		z \approx \frac{x-1}{1-3m}, \qquad  m < 1/3,
\nnn  
    		z \approx - \ln |\ln (x-1)|  \to -\infty, \qquad m=1/3.
\ear        
  For $m < 1/3$ we obtain that $W\eff  \approx - 1/(4z^2)$ as $z \to 0$;
  it is precisely the same behavior as for Fisher's solution discussed above in item 1. 
  Since the appropriate boundary condition as $z\to 0$ is here also the same, we can conclude 
  that this solution with a naked singularity in HMPG is unstable. 
  
  The case $m = 1/3$ is the most complicated. In this solution, again, $\phi \to -1$ as $x\to 1$, 
  hence the boundary condition for perturbations is
\bearr
		\delta\phi = o(1) \ \then \   \delta\psi = o\Big( (x-1)^{-1/\sqrt{3}} \Big)
\nnn
		Y = o\Big( (x-1)^{1 - 1/\sqrt{3}} \Big) \to 0.
\ear				
  On the other hand, in the limit $x \to 1$ we obtain $z \to - \infty$, more precisely,
\bearr
		z \approx - \ln |\ln (x-1)| \ \then x-1 \approx  \exp(-\e^{|z|}),
\nnn		
		W\eff \approx \frac 14 \ln^2 (x-1)  \approx \frac 14  \e^{2|z|}.
\ear
  Solving \eqn{Schr} with this asymptotic form of $W\eff$ under the assumption 
  $\omega^2 = -S^2 < 0$, we obtain a linear combination of modified Bessel functions:
\beq
		Y(z) \approx C_1 I_{S} (\e^{|z|}/2) + C_2 I_{-S} (\e^{|z|}/2).
\eeq    
   where both terms grow at large negative $z$ as $\exp(-|z|/2 + e^{|z|/2}) \to \infty$. 
   It follows that perturbations with imaginary frequencies ($\omega^2 < 0$) cannot satisfy
   the boundary condition (64), and consequently this solution is stable.

\medskip\noi
  {\bf 6.} $V(\phi) \not\equiv 0$, solution \rf{ds-q}, \rf{AE1}--\rf{UE1}, $m > 1/3$ 
  (a black hole). As follows from (39) and (60), at $m > 1/3$ there is a simple horizon 
  at some $x = x_h > 1$, where $A = 0, A' > 0$, and $W\eff = 0$, see Fig.\,7. Moreover, 
  it turns out that $W\eff < 0$ at $x > x_h$. Meanwhile, $z(x_h) = -\infty$ since the
  integral $\int dx/A(x)$ logarithmically diverges there.  It follows that $W\eff < 0$ in the 
  whole range of $z$, which inevitably leads to the existence of eigenvalues
  $E < 0$ of the quantum-mechanical eigenvalue problem with \eq (59), where it is 
  required that $Y (x)$ should be quadratically integrable.
  
  Let us determine the boundary condition for the function $Y$ in \eq (59) at $x = x_h$
  in our HMPG model. The horizon $x = x_h$ is some intermediate point, where 
  $-1 < \phi < 0$, so we have no reason to require there anything more than finiteness of
  $\delta \phi$, and since $\phi = -\tanh^2 \psi$, finiteness of $\delta\psi$ . Furthermore,
  since $Y = e^{\beta}\delta\psi$ where $\e^\beta = r(x_h)$ is finite, the boundary condition 
  at $x = x_h$ ($z \to -\infty$) is simply $|Y|  < \infty$, much weaker than would follow
  from quadratic integrability of  $Y$. It is therefore clear that the ``wave function'' 
  corresponding to $\omega^2 < 0$ as a quantum-mechanical ``energy level,'' satisfies 
  our boundary conditions and can implement instability of the black hole models under
  study.

% -------------------------------------------------------
\subsection{Stability: the phantom sector}
% -------------------------------------------------------  
  
   {\bf 1.} $V(\phi) \equiv 0$, solution (25)--(27) (the conformally mapped ``anti-Fisher'' solution).
   All branches A-C of the solution in $\ME$ contain throats $z=z_0$ (where $\beta'=0$) 
  even though not all of them correspond to \whs, and, due to $\beta'$ in the denominator,
  the potential $W\eff (z)$ for all of them contains a pole, where $W\eff (z) \approx 2/(z-z_0)^2$ .
  This singularity admits regularization by a suitable Darboux transformation, after which 
  $W\eff (z)$ is replaced by a new potential $W_{\rm reg}(z)$ that is finite and regular
  in the whole range of $u$ (or $z$) and is thus suitable for studying boundary-value problems 
  for \eq (59), as described in detail in \cite{ggs1,kb-stab11,kb-stab12,chap17}.
  
  The potential $W_{\rm reg}(z)$ has different forms for different branches of the solution
  (25)--(27). We will not present them here, referring to [41] for details. It has turned out that 
  all branches of the anti-Fisher solution are unstable [41, 51], as a result of the existence 
  of a potential well in $W_{\rm reg}(z)$. To make clear whether or not this conclusion can be
  extended to the Jordan frame (hence to HMPG), for which $W_{\rm reg}(z)$ is the same, we
  must determine the corresponding boundary conditions and compare them with those 
  applicable in the Einstein  frame.  
   
   In branches A and B ($k \geq 0$), the solution in $\ME$ exists in the range $u > 0$ that 
   corresponds to  $z \in \R$, and an unstable mode is found \cite{kb-stab11} under the 
   boundary conditions $\delta\psi \to 0$ as $z\to \pm \infty$.  However, in $\MJ$, owing to 
   the factor $\cos^2\psi$ in the metric,  the range of $u$ only extends from zero to a singular point 
   $u_s$ such that $\psi = Cu_s + \psi_0 = \pi/2$, and the range of $z$ is truncated at $z_s = z(u_s)$
   and reduces to $z \in (z_s, \infty)$. Therefore, the instability conclusion cannot be directly 
   extended to $\MJ$, and a separate new study is necessary, which is beyond the scope of this paper. 
   We can forecast that the results will depend on the solution parameters, including $\psi_0$. 
   Let us only try to formulate the appropriate boundary condition at the singularity. We
   have there $\cos \psi  = 0$, and, since it is a regular point in $\ME$, we have in its
   neighborhood 
\[    \nhq
  	\cos\psi \sim |u\!- \!u_s| \sim |z\!-\!z_s|, \quad \phi = \tan^2 \psi \sim (z\!-\!z_s)^{-2}. 
\]   
   Next, since $\phi \to \infty$ at the boundary, a reasonable condition seems to be 
   $|\delta\phi/\phi|  < \infty$, so $\delta\phi$ is allowed to behave as $(z-z_s)^{-2}$.
   But since $d\phi/d\psi = 2\sin \psi/\cos^3\psi$, we have $\delta\phi\sim \delta\psi/\cos^3\psi$,
   and our boundary condition further translates to 
\beq                  \label{z_s}
           |\delta\psi| \sim \cos\psi \sim z-z_s\ \then \  \frac{|Y|}{z-z_s} < \infty.
\eeq      
    In obtaining that, we took into account that $u_s$ is a regular point in the solution 
    in $\ME$, where, in particular, $e^\beta$ is finite, hence 
    $Y \sim \e^\beta \delta\psi \sim \delta\psi$. Thus it is the condition \rf{z_s} that should be 
    applied in the boundary-value problem for \eqn{Schr}.
    
   The same situation is found for branch C1, in the cases where a singularity also occurs due to 
   $\cos \psi = 0$. 
       
   In the wormhole case C2, the Jordan-frame solution is simply a finite deformation of its counterpart
   in $\ME$, therefore, all boundary conditions for perturbations are the same, and the instability
   conclusion from [41, 51] extends to our HMPG model.
   
   In the black hole case C3, we must formulate the condition on the horizon, which now corresponds
  to $e^\beta \sim |z| \to \infty$ in $\ME$, and simultaneously $cos  \psi \sim 1/|z| \to 0$, $\phi \to \infty$.
  Therefore, if we again require $|\delta\phi/\phi| < \infty$, we obtain then, as in (67), 
  $|\delta\psi/\cos \psi| \sim | z \delta\psi|  < \infty$. In its turn, it follows 
  $|Y| \sim \e^\beta \delta\psi \sim | z \delta\psi|  < \infty$. Thus, again, the boundary conditions in 
  $\MJ$ turn out to be less restrictive than they were in $\ME$ where the instability was established, 
  and we conclude that this result is extended to our HMPG black hole model.
      
  Lastly, in the solution (34), (35) with infinitely many horizons, any region between adjacent horizons
  is bounded by the same kind of surfaces as just discussed, with the corresponding ``weakened''
  boundary conditions, and it is straightforward to conclude that it is also unstable.       
            
\medskip\noi
 {\bf 2.} $V (\phi) \not\equiv 0$, solution (53)--(56) (the conformally mapped solution from [32] 
 describing wormholes and black universes). The following situations are possible.            
            
\medskip\noi
  (i) Solutions in which the conformal factor $\cos^2 \psi$ only deforms the Einstein-frame solution 
  in a regular manner. In these cases, all the stability results obtained for $\ME$ remain valid in $\MJ$.
  More specifically: all wormhole solutions are unstable, while among the black-universe solutions 
  there is a stable subset, in which the horizon coincides with the sphere of minimum radius (that is, 
  $x_h = 0$, where $x = x_h$ is the horizon), in all other cases the external static region $x > x_h$ 
  of a black universe is unstable [53]. These instabilities exist due to potential wells of finite depth in 
  $W_{\rm reg}(z)$, see the beginning of Subsection 5.3.

% --------------------------------------------------------------------- table 1   
\begin{table*}
\centering
\caption {HMPG solutions: Stability under monopole perturbations}
\medskip
\small
\begin{tabular}{|p{60mm}|p{80mm}| l |}
\hline
	\cm Solution           & \cm Description   & Results   \tall
\\[2pt]  \hline  \tall
		Canonical, $V \equiv 0$, (18), (19) &   
						Mapped Fisher's solution, naked singularity   & unstable
\\[2pt]
		Canonical, $V \equiv 0$, (23), $y_0 > 0$ &
						Wormhole with a conformal scalar field   & unstable
\\[2pt]
		Canonical, $V \equiv 0$, (23), $y_0 < 0$ &
						Naked singularity after a conformal continuation    & unstable
\\[2pt]
		Canonical, $V \equiv 0$, (24) &
						Black hole with a conformal scalar field   & stable
\\[2pt] \hline  \tall
		Canonical, $V \not\equiv 0$, (36)--(41), $m < 1/3$ &
						Naked singularity, similar to Fisher's    & unstable
\\[2pt]
		Canonical, $V \not\equiv 0$, (36)--(41), $m = 1/3$ &
						Naked singularity of special kind   & stable
\\[2pt]	
		Canonical, $V \not\equiv 0$, (36)--(41), $m > 1/3$ &
						Black hole with a simple horizon   & unstable
\\[2pt] \hline    \tall
		Phantom, $V \equiv 0$, (25)--(27), A, B, C1  &
						Naked singularity due to $\cos \psi  = 0$      & uncertain$^*$
\\[2pt]
		Phantom, $V \equiv 0$, (25)--(27), C2  &
						Wormhole with a phantom conformal scalar field   & unstable
\\[2pt]
		Phantom, $V \equiv 0$, (25)--(27), C3  &
						Black hole with a phantom conformal scalar field   & unstable
\\[2pt]
		Phantom, $V \equiv 0$, (34), (35)   &
						A single static region among infinitely many horizons  & unstable
\\[2pt] \hline   \tall
		Phantom, $V \not\equiv 0$, (53--(56)  &
						Naked singularity due to $\cos \psi  = 0$     & uncertain$^*$
\\[2pt]
		Phantom, $V \not\equiv 0$, (53--(56)   &
						Wormhole with an AdS or Minkowski far end   & unstable						
\\[2pt]
		Phantom, $V \not\equiv 0$, (53--(56)   &
						Black universe, generic configuration   & unstable
\\[2pt]
		Phantom, $V \not\equiv 0$, (53--(56)  &
						Black universe, horizon at minimum of $r(x)$   & stable
\\[2pt] \hline
\end{tabular}\\    
\medskip
   	${}^*$ See comments around \eqn{z_s}.
\medskip
\end{table*}                
% ------------------------------------------------------------------------------------------------

\medskip\noi         
  (ii) Solutions describing black universes in $\ME$, ``spoiled'' by a singularity $x = x_s < x_h$ due
  to $\cos \psi  = 0$, i.e., there is a big-bang-like singularity located in the T-region beyond the horizon. 
  The stability results for the external region $x > x_h$ remain the same as in item (i).
          
\medskip\noi               
   (iii) Solutions with naked singularities $x = x_s$ in a static region due to $\cos \psi  = 0$, which are
   possible at any value of $x$ in wormhole solution or with any $x_s > x_h$ in black-universe solutions 
   in $\ME$.  In all such cases, the situation looks the same as previously discussed for $V \equiv 0$: 
   we have a truncated range $x_s < x < \infty$, with the boundary condition (67) at $x = x_s$, 
   and a separate study is necessary to find out the exact (in)stability conditions. 

\medskip\noi           
   The stability results are summarized in Table 1.       
                                                            
% ================================
 \section{Concluding remarks}
% ================================
      
  We have considered exact analytical vacuum static, spherically symmetric asymptotically 
  flat solutions of HMPG, using its scalar-tensor representation, with both zero and 
  nonzero potentials $V (\phi)$, on the basis of known solutions of GR with minimally
  and conformally coupled scalar fields. All configurations split into two large classes, 
  one corresponding to a canonical scalar field ($-1 < \phi < 0$), the other to a phantom 
  one ($\phi > 0$).
  
  It has been stated that in the case $V \equiv 0$ most of the HMPG space-times contain 
  naked singularities, and a generic family of solutions in the phantom sector, as could 
  be expected, describes traversable wormholes. As to possible black holes,
  it turns out that that there are only two special families (one in the canonical sector 
  and another in the phantom one) that describe extremal black holes (hence having 
  zero Hawking temperature), and the one with a phantom scalar is globally regular.
  Such results substantially disagree with those of [14], where the same problem was 
  studied numerically with equations written in the in Jordan frame, and black hole
  solutions with simple (finite-temperature) horizons were found. The reason for
  this disagreement is yet to be understood.  
  
  To obtain examples of exact solutions with $V \not\equiv 0$, we have used the 
  previously obtained solutions of GR in which black hole subsets (this time with simple 
  horizons) are generic. Naturally, in the canonical sector this can only happen
   with at least partly negative potential $V(\phi)$ since the well-known no-hair theorem 
   from GR [47] (on nonexistence of black holes with variable minimally coupled scalar 
   fields with nonnegative potentials) directly extends to the Jordan frame as long as
   the corresponding conformal factor is well-behaved. Here we again disagree with 
   [14] where a number of HMPG black hole solutions were obtained numerically,
  and some of them with $V > 0$. Further studies are probably necessary in order 
  to explain this contradiction.

  In the phantom sector, generic black hole solutions are of black-universe type [32, 33], 
  there are also wormholes with  flat or AdS asymptotics at the far end. And, with both
  zero and nonzero potentials, there emerge a new kind of singularities due to vanishing 
  of the conformal factor $\cos^2\psi$; depending on the solution parameters, such 
  singularities may be located in a static region (it is then a singular attracting center) 
  or beyond a black hole horizon (it is then like a big bang or big crunch).     
  
  It has turned out that most of the solutions under study are unstable under spherically
  symmetric monopole perturbations. Some of these instability results have been 
  extended from their counterparts known in GR (but certainly taking into account
  the boundary conditions formulated in the Jordan frame), some others have been 
  obtained anew, see their summary in Table 1. Only some special solutions prove 
  to be stable, including the well-known black hole with a massless conformal       
  scalar field [24, 25, 60] and a conformally mapped black universe with a horizon 
  at the minimum radius [32, 53]. 
  
  In conclusion, let us mention some possible directions of continuation or extensions 
  of the present study. First of all, in the case of zero potential ($V =0$) it is straightforward 
  to obtain similar solutions with electromagnetic fields $F\mn$, by analogy with 
  previous studies in scalar-tensor theories [23, 29]. With nonzero potentials, similar
  configurations with electromagnetic fields can also be treated both analytically and 
  numerically, e.g., on the basis of known GR solutions [44, 54]. Another trend of interest 
  is a consideration of similar problems in the so-called extended HMPG containing
  functions $f(R, \cR)$ of two curvatures [9, 10, 61], whose scalar-tensor representation 
  contains two interacting scalar fields.
    
% ------------------------------------------------------------      
\subsection*{Acknowledgments}

  The work was funded by the RUDN University Program 5-100 and the Russian Basic 
  Research Foundation grant 19-02-0346. The work of K.B. was also partly performed 
  within the framework of the Center FRPP supported by MEPhI Academic
  Excellence Project (contract No. 02.a03.21.0005, 27.08.2013).

% ----------------------------------------------------------------------------------

% =============================================
\end{document}